\documentclass[manuscript,screen,table]{acmart}

\renewcommand\footnotetextcopyrightpermission[1]{} 
\AtBeginDocument{%
  }

\setcopyright{acmlicensed}
\copyrightyear{2018}
\acmYear{2018}
\acmDOI{XXXXXXX.XXXXXXX}
\acmConference[Conference acronym 'XX]{Make sure to enter the correct
  conference title from your rights confirmation email}{June 03--05,
  2018}{Woodstock, NY}
\acmISBN{978-1-4503-XXXX-X/2018/06}

\usepackage[utf8]{inputenc}
\usepackage{amsmath}
\usepackage{enumerate}
\usepackage{threeparttable}
\usepackage{booktabs} 
\usepackage{graphicx}
\usepackage{calc}
\usepackage{xcolor}
\usepackage{color}
\definecolor{light-gray}{gray}{0.80}
\definecolor{grannysmithapple}{rgb}{0.66, 0.89, 0.63}
\definecolor{green(html/cssgreen)}{rgb}{0.0, 0.5, 0.0}
\definecolor{brightmaroon}{rgb}{0.76, 0.13, 0.28}

\usepackage{framed}
\usepackage{amsmath}

\newlength{\DepthReference}
\newlength{\HeightReference}
\newlength{\Width}%

\newcommand{\MyColorBox}[2][red]%
{%
    \settowidth{\Width}{#2}%
    \colorbox{#1}%
    {%
        \raisebox{-\DepthReference}%
        {%
                \parbox[b][\HeightReference+\DepthReference][c]{\Width}{\centering#2}%
        }%
    }%
}

\usepackage{pifont}
\usepackage{arydshln} 
\usepackage{listings}
\usepackage{lipsum}
\lstdefinestyle{interfaces}{
  float=tp,
  floatplacement=tbp,
  abovecaptionskip=0pt,
} 

\definecolor{codegreen}{rgb}{0,0.6,0}  
\lstdefinelanguage{diff}{  
  morecomment=[f][\color{blue}]{@@},     
  morecomment=[f][\color{red}]-,         
  morecomment=[f][\color{codegreen}]+,       
  morecomment=[f][\color{red}]{---}, 
  morecomment=[f][\color{codegreen}]{+++},
}

\usepackage{graphicx}

\usepackage{subcaption}
\usepackage{cleveref}

\usepackage{enumitem} 
\usepackage{booktabs}
\usepackage{multirow}
\usepackage{makecell}

\usepackage{pgfplots}

\usepackage{colortbl} 
\usepackage{rotating}

\usepackage{enumitem}

\begin{document}

\title{Towards Reliable LLM-Driven Fuzz Testing: Vision and Road Ahead}

\author{Yiran Cheng}
\affiliation{%
 \institution{Beijing Key Laboratory of IOT Information Security Technology, Institute of Information Engineering, CAS; School of Cyber Security, University of Chinese Academy of Sciences}
 \city{Beijing}
 \country{China}}
\email{chengyiran@iie.ac.cn}

\author{Hong Jin Kang}
\affiliation{%
 \institution{University of Sydney}
 \city{Sydney}
 \country{Australia}}
\email{hongjin.kang@sydney.edu.au}

\author{Lwin Khin Shar}
\affiliation{%
 \institution{Singapore Management University}
 \city{Singapore}
 \country{Singapore}}
\email{lkshar@smu.edu.sg}

\author{Chaopeng Dong}
\affiliation{%
 \institution{Beijing Key Laboratory of IOT Information Security Technology, Institute of Information Engineering, CAS; School of Cyber Security, University of Chinese Academy of Sciences}
 \city{Beijing}
 \country{China}}
\email{dongchaopeng@iie.ac.cn}

\author{Zhiqiang Shi}
\affiliation{%
 \institution{Beijing Key Laboratory of IOT Information Security Technology, Institute of Information Engineering, CAS; School of Cyber Security, University of Chinese Academy of Sciences}
 \city{Beijing}
 \country{China}}
\email{shizhiqiang@iie.ac.cn}

\author{Shichao Lv}
\affiliation{%
 \institution{Beijing Key Laboratory of IOT Information Security Technology, Institute of Information Engineering, CAS; School of Cyber Security, University of Chinese Academy of Sciences}
 \city{Beijing}
 \country{China}}
\email{lvshichao@iie.ac.cn}

\author{Limin Sun}
\affiliation{%
 \institution{Beijing Key Laboratory of IOT Information Security Technology, Institute of Information Engineering, CAS; School of Cyber Security, University of Chinese Academy of Sciences}
 \city{Beijing}
 \country{China}}
\email{sunlimin@iie.ac.cn}

\renewcommand{\shortauthors}{Cheng et al.}

\begin{abstract}
  Fuzz testing is a crucial component of software security assessment, yet its effectiveness heavily relies on valid fuzz drivers and diverse seed inputs. 
  Recent advancements in Large Language Models (LLMs) offer transformative potential for automating fuzz testing (LLM4Fuzz), particularly in generating drivers and seeds. 
  However, current LLM4Fuzz solutions face critical reliability challenges, including low driver validity rates and seed quality trade-offs, hindering their practical adoption. 
  
  This paper aims to examine the reliability bottlenecks of LLM-driven fuzzing and explores potential research directions to address these limitations.
  It begins with an overview of the current development of LLM4SE and emphasizes the necessity for developing reliable LLM4Fuzz solutions. Following this, the paper envisions a vision where reliable LLM4Fuzz transforms the landscape of software testing and security for industry, software development practitioners, and economic accessibility. 
  It then outlines a road ahead for future research, identifying key challenges and offering specific suggestions for the researchers to consider. 
  This work strives to spark innovation in the field, positioning reliable LLM4Fuzz as a fundamental component of modern software testing.
\end{abstract}


\maketitle

\section{Introduction}
\begin{figure}
    \centering
    \includegraphics[width=0.65\linewidth]{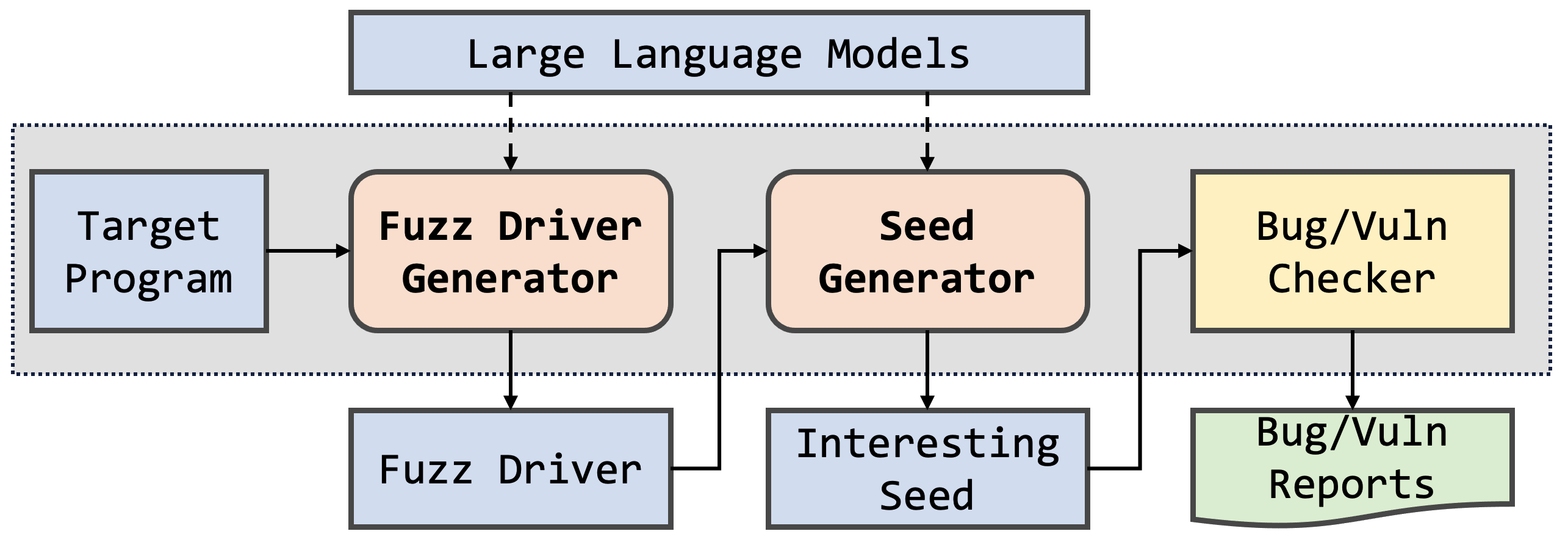}
    \caption{The Architecture of LLM4Fuzz Framework.}
    \label{fig:workflow}
\end{figure}

Fuzz testing is a fundamental technique in software security assessment, which continuously feeds abnormal or random inputs to target programs to discover potential vulnerabilities and defects~\cite{zhang2023bg,chafjiri2024bg}. The effectiveness of fuzzing largely depends on two critical components as demonstrated in Figure~\ref{fig:workflow}: fuzz drivers that interface with the target program's APIs~\cite{manes2019fuzzdriverintro,liang2018fuzzdriverintro2}, and high-quality seeds that trigger diverse program behaviors~\cite{li2018seedintro,wang2017seedintro}. The fuzz drivers feed the seeds into the target program to trigger bugs or vulnerabilities. 

Recent advances in Large Language Models (LLMs) have demonstrated remarkable capabilities in code understanding and generation tasks~\cite{ahmed2024task,du2024task,fan2023programrepair,hossain2024programrepair}. LLMs like GPT-4~\cite{achiam2023gpt4} and CodeLlama~\cite{roziere2023codellama} pre-trained on vast amounts of code and natural language data have shown potential in addressing various software engineering challenges. 
Given their success in code-related tasks, researchers have begun exploring the application of LLMs to enhance automated fuzz testing (LLM4Fuzz). 

Nowadays, LLM4Fuzz offers promising opportunities in two key areas: (1) automating the generation of fuzz drivers by learning from API documentation and usage patterns~\cite{CKGFuzzer,ma2024mGPTFuzz,zhang2024howeffective,deng2024edgecase,TitanFuzz,PromptFuzz}, and (2) producing diverse and semantically seed values through their understanding of program behaviors and input formats~\cite{xia2024fuzz4all,oliinyk2024fuzzingbusybox,eom2024fuzzjs,meng2024protocolfuzz,yu2024llmfuzzer,zhou2024magneto,wang2024prophetfuzz,fu2024sedar,zhao2023understanding,wang2024llmif}. 
These approaches have brought new levels of automation to security testing, significantly improving testing coverage and vulnerability detection capabilities. 
However, due to their cost-effectiveness concerns, most existing LLM4Fuzz approaches are predominantly explored in research environments. 
The broader security testing community, including small businesses and individual developers, faces significant barriers to adopting these technologies. 
Before LLM4Fuzz solutions can become widely accessible, several key challenges need to be addressed, most notably their \emph{reliability}, including the usability of generated drivers and syntactic and semantic validity of seeds. 

\noindent\textbf{Necessity for Reliable LLM4Fuzz.}
Reliable LLM4Fuzz involves generating valid fuzz drivers and high-quality (interesting) seeds that can test target programs without false positives. To date, developing LLM4Fuzz solutions faces reliability challenges, with many generated outputs failing basic validation checks. For example, FuzzGPT~\cite{deng2024edgecase} demonstrates limited valid driver generation rates, achieving only 27.43\% and 17.41\% success rates on PyTorch~\cite{Pytorch} and TensorFlow~\cite{TensorFlow} respectively. 
This unreliability stems from various factors, including incorrect API usage patterns, invalid parameter combinations, and improper handling of dependencies.
Generated seeds face similar challenges, with Fuzz4All~\cite{xia2024fuzz4all} showing validity rates 56\% lower than traditional methods, leading to ineffective testing cycles and missed vulnerabilities. 
Such reliability issues often force practitioners to implement extensive validation mechanisms and manual verification steps, significantly increasing the overall testing time and resource consumption~\cite{babic2019fudge, takanen2018fuzzing}. 
Consequently, security testing efforts using unreliable LLM4Fuzz solutions may miss critical vulnerabilities or waste resources on invalid test cases. Building reliable LLM4Fuzz solutions is therefore crucial for making these technologies practical and trustworthy for real-world security testing applications.


\noindent\textbf{Challenges and Opportunities.} 
In this paper, we identify four common challenges that critically influence the reliability of LLM4Fuzz. C1: complex API dependencies hinder accurate modeling of multi-layered initialization sequences and domain-specific constraints. C2: the lack of driver quality assurance results in frequent runtime failures despite compilation checks. C3: seed quality trade-offs arise from the tension between maintaining input validity and exploring diverse program paths. C4: ensuring semantic consistency between generated seeds and program requirements remains elusive, particularly for format-specific inputs. These challenges collectively undermine the practicality of LLM4Fuzz in real-world security testing. 

To address these limitations, we propose targeted opportunities to enhance LLM-assisted reliability across fuzzing steps—from driver synthesis to seed validation—while balancing automation and robustness. 
By exploring these opportunities, we quantify the reliability through a set of metrics, enabling more practical and scalable LLM4Fuzz solutions for diverse testing scenarios.
\section{Current Development}
This section presents the current state of LLM4Fuzz solutions regarding reliability for fuzz driver and seed generation. Here, we group the literature based on the key techniques used.

\subsection{Driver Generation}
\begin{figure*}[t]
    \centering
    \includegraphics[width=0.98\linewidth]{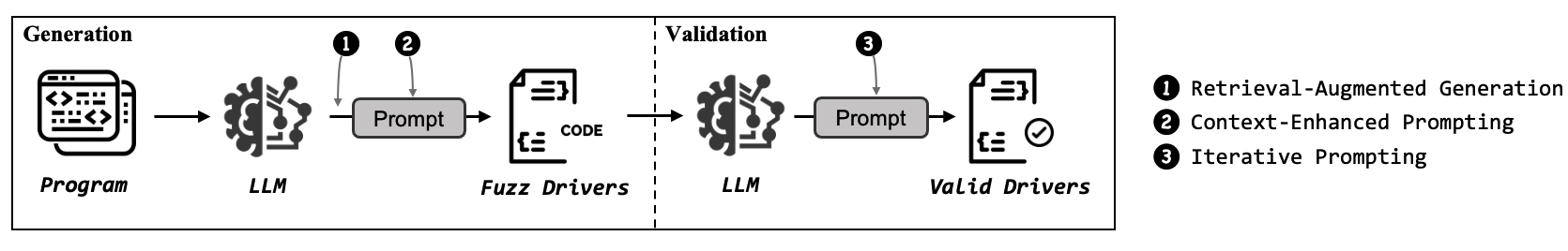}
    \caption{LLM-Driven Fuzz Driver Generation Pipeline.}
    \label{fig:driver_pipeline}
\end{figure*}

The basic idea is to use the target programs as the prompt context, and then ask LLMs to generate API sequences as fuzzing drivers, such as ``\textit{The following is a fuzz driver written in C language, complete the implementation}''~\cite{zhang2024howeffective}. To improve the reliability of the generated driver, many approaches employ adaptation strategies for directing generation in different steps, shown in Figure~\ref{fig:driver_pipeline}. 

\noindent\textbf{External Knowledge Infusion.}
Retrieval-Augmented Generation (RAG \ding{182}) is a technique that combines a retrieval system to fetch relevant information from external knowledge sources with the language model's generation capabilities~\cite{lewis2020rag1,gao2023rag2}. 
CKGFuzzer~\cite{CKGFuzzer} employs RAG where the external knowledge comes from a code knowledge graph containing API implementations, function call relationships, and API contextual to improve the reliability of generated drivers. 
Taking a target API as a query index, CKGFuzzer retrieves the relevant API information from the code knowledge graph, ranks APIs' relevance to the query, and then generates the final API combinations.

\noindent\textbf{Context-aware Prompting.}
Zhang et al.~\cite{zhang2024howeffective} extends the basic strategy by incorporating fundamental API context in the prompts (\ding{183}). It extracts API declarations from the header file's Abstract Syntax Tree (AST), which includes both function signatures and argument variable names. For example, when generating a fuzz driver for the \texttt{bpf\_open\_mem} function, the prompt would include the header file \texttt{bpf/libbpf.h}, the complete function declaration "\texttt{extern bpf* bpf\_open\_mem(char *buf, int sz, struct opts *opts)}", and a request to implement the fuzzing function.

\noindent\textbf{Generate-and-fix Validation.}
The iterative prompting strategy (\ding{184}) is employed in Zhang et al.'s work~\cite{zhang2024effective}. 
They follow a generate-and-fix workflow where the model first attempts to generate a fuzz driver and then enters a refinement cycle if the initial generation fails validation. In the refinement process, the system collects error information such as compilation errors and constructs fix prompts. Each fix prompt includes the erroneous driver code, error summary, problematic code lines, and supplementary API information when available. This iterative process continues until either a valid driver is produced or the maximum iteration limit (5 rounds) is reached.

\subsection{Seed Generation}
\begin{figure*}[t]
    \centering
    \includegraphics[width=1.0\linewidth]{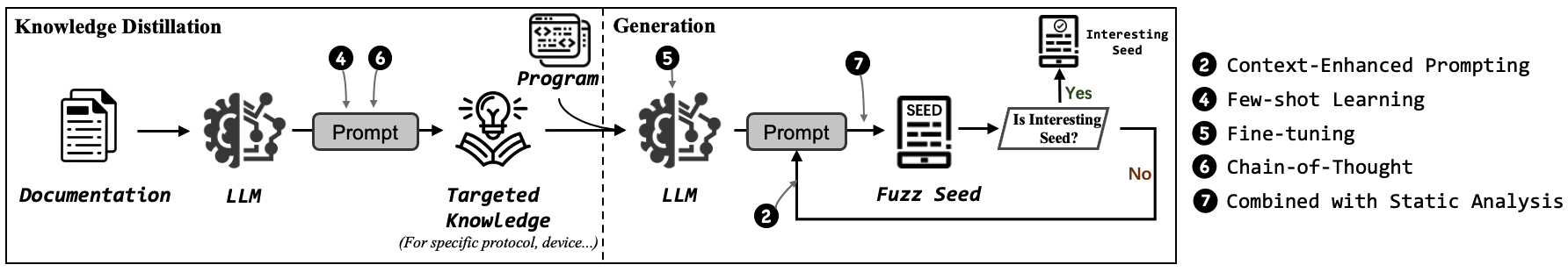}
    \caption{LLM-Driven Seed Generation Pipeline.}
    \label{fig:seed_pipeline}
\end{figure*}

The basic idea is to use the target program as the prompt context and then ask LLMs to generate new seeds. For example, Asmita et al.~\cite{oliinyk2024fuzzingbusybox} use the following prompt: \textit{"Generate initial seed to fuzz BusyBox awk applet"}. To improve the reliability of generated seed, many approaches employ different strategies to direct knowledge distillation and generation steps, shown in Figure~\ref{fig:seed_pipeline}. 

\noindent\textbf{Knowledge Distillation Prompting.}
Some approaches conduct a knowledge distillation step by helping LLMs correctly understand documentation information to provide targeted knowledge for seed generation.
To address the challenge of vast option combination spaces from the program manpage that potentially harbor risks, ProphetFuzz~\cite{wang2024prophetfuzz} employs eight examples (few-shot learning \ding{185}) to guide GPT-4 in identifying dangerous option combinations from manpage, training the LLM to explore vulnerable option combinations. 
Before conducting IoT protocol fuzzing, LLMIF~\cite{wang2024llmif} employs CoT prompting (\ding{187}) to analyze dependencies between device messages. It guides the model through a reasoning process to understand how the first and second messages interact with device properties. The identified dependencies are then utilized to guide subsequent seed mutation.

\noindent\textbf{Combination with Other Models.}
Standalone LLMs have some limitations in generating valid and interesting seeds: 1) 
when LLMs are used with generic prompts such as  "\textit{Please mutate the following program}", they tend to generate large blocks of code; 2) LLMs tend to make conservative changes, which is not conducive to exploring new program paths. 
Eom et al.~\cite{eom2024fuzzjs} combined Masked Language Modeling with LLM (\ding{186}) to address the limitations. They adopted three masking strategies: Insert (inserting [MASK] tokens at random positions), Overwrite (replacing existing tokens with [MASK]), and Splice (dividing code into segments and replacing them with [MASK]-tagged segments from other seeds). When a masked sequence is created, it is fed into the CovRL-trained LLM mutator for inference, allowing it to learn which positions of mutation are more likely to improve coverage.

\noindent\textbf{Combination with Static Analysis (\ding{188}).}
CHATAFL~\cite{meng2024protocolfuzz} is a LLM-based protocol fuzzer combined with grammar-guided mutation (static analysis). Grammar-guided mutation aims to perform structure-aware mutations by leveraging protocol grammar information extracted from LLMs, which can improve the reliability of mutated seeds. During the mutation process, the fuzzer only modifies mutable fields while preserving the overall message structure. For example, in an \texttt{RTSP PLAY} request, the tool modifies the values of \texttt{CSeq} or \texttt{Range} fields but maintains the overall message format. This grammar-guided mutation ensures that the generated test inputs are structurally valid.

\noindent\textbf{Enhancement for Interesting Seed Generation.}
CHATAFL~\cite{meng2024protocolfuzz} employs context-enhanced prompting (\ding{183}) when facing coverage plateaus during fuzzing. When detecting fuzzer generates consecutive uninteresting messages (i.e., seeds that do not increase coverage), it collects the recent communication history between the client and server and incorporates this context into the LLM prompts. This contextual information enables the LLM to understand the current protocol state and generate messages (seeds) that are likely to trigger transitions to new states.

\section{Vision}
This section envisions a future where reliable LLM4Fuzz revolutionizes security testing in software engineering. We explore this vision through the perspectives of industry, software practitioners, and society, illustrating how LLM4Fuzz will transform vulnerability discovery, secure development practices, and the overall security landscape.

\noindent\textbf{For Industry.}
Advances in reliable LLM4Fuzz will facilitate the development of \emph{fast yet trustworthy security testing solutions} that balance speed and reliability. The existing security testing landscape is dominated by heavyweight fuzzing tools (e.g., AFL++~\cite{fioraldi2020afl++}, libFuzzer~\cite{LibFuzzer}) that require massive computational resources and prolonged execution time. For instance, Google's OSS-Fuzz~\cite{serebryany2017ossfuzz} spends over 50 CPU hours to generate a single Node.js driver~\cite{jeong2023utopia}, while industrial-grade fuzzers often demand days or weeks to achieve meaningful coverage. These inefficiencies lead to delayed vulnerability discovery and inflated operational costs.

By adopting reliable LLM4Fuzz, companies can replace such resource-intensive workflows with lightweight alternatives that ensure both reliability and speed. 
These solutions seamlessly integrate into standard development infrastructure, enabling real-time vulnerability detection without compromising accuracy. Consequently, organizations can achieve faster security testing cycles (e.g., completing full-coverage tests in hours instead of days) while ensuring the reliability of generated test cases, ultimately minimizing risks and costs.

\noindent\textbf{For Software Development Practitioners.}
Advances in reliable LLM4Fuzz will pave the way for the emergence of \emph{intelligent security companions that work alongside developers in their secure development journey}. Currently, developers rely on cloud-based security testing services (i.e., Synopsys Coverity~\cite{coverity}, Checkmarx SAST~\cite{Checkmarx}, or Mayhem Security~\cite{Mayhem}) that pose potential privacy risks~\cite{amil2024impact,chatterjee2015security}. 
Many developers are hesitant to use these services as they require sharing sensitive code with third parties, while existing local tools lack customization options and the intelligence to understand project-specific security requirements and development patterns.

By adopting LLM4Fuzz, practitioners will gain access to dedicated security assistants that operate privately within their local development environments, protecting their intellectual property and sensitive code. These companions will adapt to individual development styles and project-specific security contexts, evolving their testing strategies based on the unique characteristics of each project and developer's preferences. The assistants will offer immediate, interactive security feedback while maintaining complete privacy, enabling developers to understand and verify potential vulnerabilities as they code. 
This symbiotic relationship between developers and their security companions will transform security testing from a separate activity into an integral part of the development workflow.

\noindent\textbf{Economic Accessibility.}
Advances in reliable LLM4Fuzz will drive the development of \emph{accessible security testing infrastructure for the digital ecosystem}. The current state of software security testing is often restricted to well-funded organizations due to the high computational and operational costs involved. Many smaller companies struggle to implement adequate security testing due to resource constraints~\cite{tam2021smallbs,kabanda2018smallbs}. Meanwhile, commercial security testing services require subscriptions that can cost tens to hundreds of thousands of dollars depending on the codebase size and testing frequency. For example, Gitlab reports Coverity costs around 12k USD per year~\cite{Gitlab_report}.

By adopting reliable LLM4Fuzz solutions, we can democratize advanced security testing capabilities through lower resource requirements and operational costs. This widespread adoption will enable more organizations to implement thorough security testing, leading to earlier detection and remediation of vulnerabilities across the software ecosystem. Furthermore, the reduced resource requirements will decrease the carbon footprint of security testing operations. This aligns well with the growing societal emphasis on environmental sustainability, allowing organizations to meet their security needs while contributing to carbon reduction goals.
\section{The Road Ahead}
\begin{figure}
    \centering
    \includegraphics[width=0.75\linewidth]{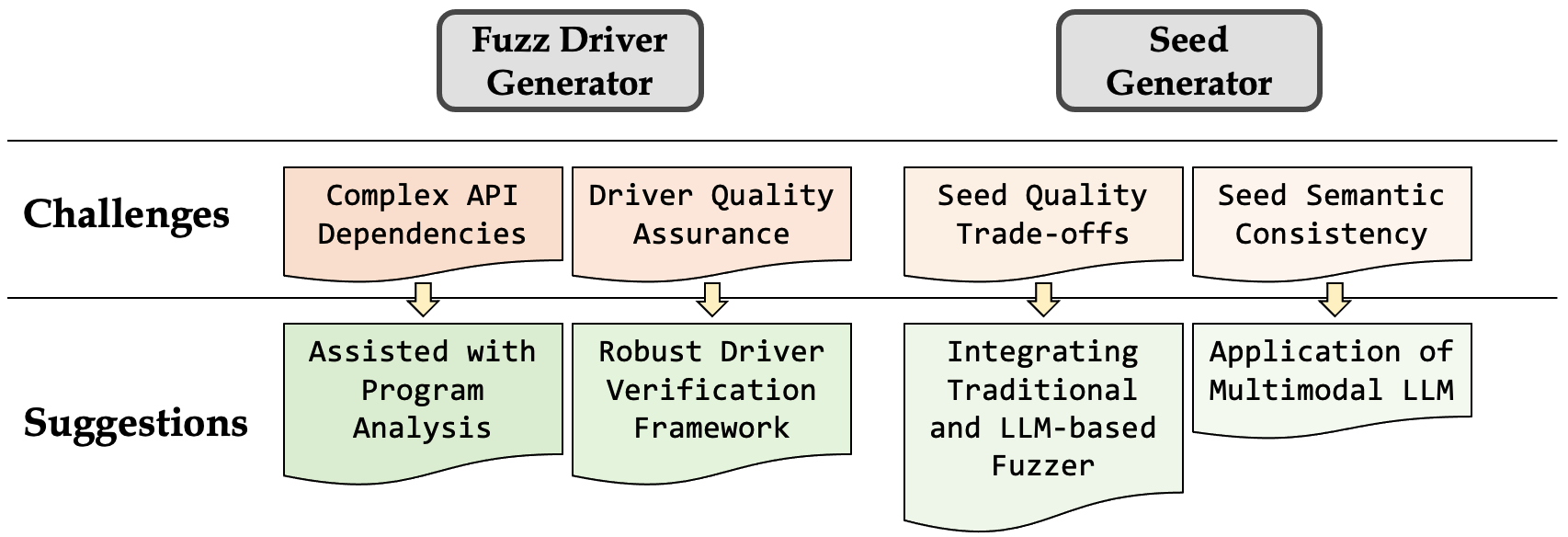}
    \caption{Challenges and Future Research Suggestions for Reliable LLM4Fuzz.}
    \label{fig:roadmap}
\end{figure}

In this section, we discuss current challenges (Section 4.1) and corresponding future research suggestions (Section 4.2) for enhancing reliable LLM4Fuzz solutions, as demonstrated in Figure~\ref{fig:roadmap}. By exploring these directions, we aim to quantify the reliability through a set of metrics (Section 4.3), paving the way for more reliable LLM4Fuzz solutions that can meet the growing demands of modern software security.

\subsection{Challenges}

\noindent\textbf{C1: Complex API Dependencies.} 
The application of LLMs to reliable driver generation faces fundamental challenges in capturing complex API relationships. 
Although LLMs have learned the basic pattern of driver code and common programming from training in vast amounts of data, the multi-layered nature of API dependencies creates intricate initialization sequences and conditional constraints that are difficult to accurately model~\cite{zhang2023open}. 
Additionally, domain-specific API constraints and business logic often require specialized knowledge that may not be adequately represented in LLMs' training data.
This is evidenced in Zhang et al.'s work~\cite{zhang2024howeffective} where LLMs failed to resolve 6\% (5/86) of questions due to complex API dependency requirements, such as creating standby network servers/clients or constructing proper relationships between various concepts like sessions and windows in \texttt{tmux}.

\noindent\textbf{C2: Driver Quality Assurance.} 
While certain approaches offer some quality assurance such as compilation error verification and runtime check, the reliability of generated drivers remains a critical concern. Current approaches~\cite{CKGFuzzer,zhang2024howeffective,TitanFuzz} lack robust formal verification mechanisms for ensuring the correct API usage in generated drivers, for example, the verification for API call sequence and API parameter constraints. While semantic checkers can verify correctness, their development requires significant effort, limiting practical application.
As evidenced by the limited effective program generation rates in FuzzGPT~\cite{deng2024edgecase} (only 27.43\% and 17.41\% on PyTorch and TensorFlow respectively), a significant portion of generated driver code fail to pass basic runtime checks. A more robust formal verification mechanism could potentially help identify incorrect API usage early, thereby improving the validity rate of generated programs.

\noindent\textbf{C3: Seed Quality Trade-offs.}
The generation of high-quality seeds presents multiple competing objectives. Despite achieving a 36.8\% increase in code coverage, Fuzz4All~\cite{xia2024fuzz4all} shows that LLM-generated seeds demonstrate significantly lower validity rates, averaging 56\% below traditional methods. Unlike traditional fuzzers that are specifically designed with built-in language rules and constraints, LLM is a general-purpose model without such specialized constraint mechanisms.
While striving to improve validity, maintaining sufficient diversity becomes particularly challenging. CovRL-Fuzz~\cite{eom2024fuzzjs} found that LLM-based mutations tend to conduct grammar-level mutations, which target grammatical accuracy but also limit variability.

\noindent\textbf{C4: Semantic Consistency.}
Ensuring semantic alignment between generated seeds and program requirements poses a substantial challenge. For instance, as observed in prior work~\cite{wang2024llmif}, when dealing with format-specific options like "-f avi", the generated file must strictly conform to the AVI format specifications. This requires not only syntactic correctness but also proper handling of format-specific structures like headers, chunks, and index tables.  However, existing approaches ignore the need to address how to ensure format consistency.

\subsection{Future Research Suggestions}

\noindent\textbf{S1: API Knowledge Enhancement Assisted with Program Analysis.}
Knowledge enhancement through static and dynamic program analysis (e.g., symbolic execution, data flow analysis) provides a promising avenue for improving context-aware driver synthesis. 
By applying interprocedural analysis on target interfaces and existing test programs, we can automatically extract API usage patterns (e.g., parameter dependency chains, state transition sequences) and data dependency relationships. This enables constraint-aware driver generation that respects target-specific input validation logic and execution constraints.

The framework can be strengthened by developing hybrid code representation models combining control-flow-sensitive abstractions (e.g., call graphs with path conditions) with data-flow annotations. This facilitates the construction of domain-specific API knowledge graphs that encode: 1) Parameter type and value constraints 2) State transition dependencies 3) Error handling patterns. For network protocol fuzz drivers, domain modeling should explicitly capture protocol state machines, message format specifications, and timing constraints through formal specification languages like P4 or Statecharts.


\noindent\textbf{S2: Robust Driver Verification Framework.} 
To enhance the reliability of generated drivers, the system should employ robust verification frameworks that integrate static analysis (e.g., code structure and API call sequence validation) and dynamic analysis (e.g., runtime behavior and parameter constraint checks). Formal methods, such as model checking or theorem proving, can be applied to rigorously validate both the correctness of API call sequences and the adherence to parameter constraints. This addresses a key limitation in current approaches like FuzzGPT, where generated drivers often fail basic runtime checks due to insufficient validation.

Further, the system could include automated semantic checker generation tools that parse API documentation (e.g., function descriptions, parameter types, and constraints) to create tailored semantic validators. This would significantly reduce the manual effort required for semantic validation, streamlining the process.

\noindent\textbf{S3: Hybrid Method Integrating Traditional Fuzzer and LLM-based Fuzzer.}
Traditional fuzzers excel at maintaining strict validity by leveraging built-in grammar rules and input format constraints, which ensure that generated test cases conform to the expected structure of the target program. However, these fuzzers often struggle to explore deeper semantic behaviors due to their reliance on random or rule-based mutations. On the other hand, large language models (LLMs) demonstrate superior code coverage capabilities by generating diverse and semantically rich input variations, albeit with lower validity rates. By integrating these complementary strengths, we can develop a more robust testing strategy. 

Specifically, LLMs can enhance traditional fuzzing by providing a deeper understanding of input formats and program logic, enabling the generation of more meaningful mutations that push the program into unexplored states. Meanwhile, traditional fuzzers can enforce validity constraints on these LLM-generated seeds, ensuring they adhere to the required syntax and structural rules. This hybrid approach addresses the limitations observed in prior works~\cite{xia2024fuzz4all,eom2024fuzzjs}, where LLMs often produced invalid inputs or lacked diversity in mutation strategies. By combining the semantic richness of LLMs with the strict validity guarantees of traditional fuzzers, the hybrid method aims to achieve both high code coverage and robust input validity.

\noindent\textbf{S4: Application of Multimodal LLM.}
While existing approaches struggle with format consistency~\cite{grieco2016quickfuzz,fell2017review}, as demonstrated in prior work where generating format-specific files (e.g., AVI) proved challenging due to issues like misaligned headers or incomplete index tables~\cite{wang2024llmif}, multimodal LLMs emerge as a promising solution to address semantic alignment in seed generation. These models' capability to process and learn from diverse data types—such as images, videos, and audio—enables them to internalize not only the syntactic specifications but also the structural intricacies and constraints specific to each format~\cite{mckinzie2025mm1,hu2024bliva}. 

For example, when tasked with generating an AVI file, a multimodal LLM could leverage its understanding of the format's mandatory components (e.g., chunk layout, header fields) to ensure the output adheres to the required specifications. By learning from real-world examples, these models can generate seeds that are both syntactically valid and semantically consistent with the program's requirements, directly producing files in the target format without intermediate conversions or manual adjustments.

\subsection{Reliability Assessment Metrics}
To ensure the practical adoption of LLM4Fuzz, it is crucial to define and quantify its reliability through a set of well-defined metrics. This section proposes metrics for evaluating driver validity and seed quality, which are essential for assessing the effectiveness of LLM4Fuzz.

\noindent\textbf{Driver Validity.}
To evaluate the validity of the generated drivers, we conclude three metrics: 1) Compilation success rate: the percentage of generated fuzz drivers that successfully compile without errors.
2) Runtime success rate: the percentage of compiled drivers that execute without runtime failures (e.g., crashes or exceptions).
3) API usage correctness: the adherence of generated drivers to API usage patterns, including parameter constraints and call sequences.

\noindent\textbf{Seed Quality.}
To evaluate the validity of the generated seeds, we conclude three metrics: 
1) Validity rate: the percentage of generated seeds that conform to the syntactic and semantic requirements of the target program.
2) Path coverage: the number of unique program paths triggered by the seeds, measured using code coverage tools (e.g., gcov, LLVM Coverage).
3) Vulnerability discovery rate: the number of unique vulnerabilities discovered by the seeds, categorized by severity (e.g., critical, high, medium, low).

These metrics are designed to comprehensively evaluate the correctness of generated artifacts (drivers and seeds) and their effectiveness in achieving the core objectives of fuzz testing—triggering diverse program behaviors and discovering crashes and vulnerabilities.


\section{Conclusion and Future Work}
In this paper, we summarize the current development in exploring the potential of reliable LLM4Fuzz. Looking ahead, we envision LLM4Fuzz evolving into a transformative force in security testing - making it more accessible for industry, providing developers with intelligent security companions, and democratizing testing infrastructure for broader societal benefit. 
To outline the road ahead, We highlight four critical challenges: complex API dependencies, insufficient driver quality assurance, seed quality trade-offs, and semantic consistency issues. To address these challenges, we propose several research suggestions, which focus on creating standardized reliability benchmarks, optimizing seed generation, and exploring multimodal LLM applications. As LLM technology advances, addressing these challenges while pursuing suggested research directions will lead to more reliable security testing tools.


\end{document}